# Long-range correlation in the whole human genome

R. Mansilla, N. Del Castillo, T. Govezensky, P. Miramontes, M. José and G. Cocho

1. **Abstract**

We calculate the mutual information function for each of the 24 chromosomes in the human genome. The same correlation pattern is observed regardless the individual functional features of each chromosome. Moreover, correlations of different scale length are detected depicting a multifractal scenario. This fact suggest a unique mechanism of structural evolution. We propose that such a mechanism could be an expansion-modification dynamical system.

# I Introduction

The sequencing of the human genome is probably the most important episode in nucleic acids biology after the disclosure of the DNA double helix structure by J.D. Watson and F.H.C. Crick. The availability of the human genome has opened new fields in medicine, forensic sciences and anthropology. Nonetheless, the relationship between functional and structural traits is a rather unexplored field.

It is well known that human chromosomes are very different among themselves. Putting aside the obvious differences in size, there are also divergences in the density and spatial localization of genes and organization of Alu repeats [1], the distribution of CpG islands[2], etcetera. For instance, Chromosome 19 has 57% of repetitive elements whilst chromosomes 2, 8, 10, 13 and 18 have 36%. The chromosome 19 also has the highest gene density (around 29%, depending on the method) against 10% in chromosomes 4, 13 and 18.

Many efforts have been made to unveil the nature of genome evolution. Gene duplication is a well accepted mechanism but it cannot explain the evolution of non-coding DNA. The challenge is then to find an evolutionary mechanism compatible with both the protein coding requirements and the structure of non-coding DNA.

The aim of this paper is to carry out an study of the long-range correlation properties of the complete human genome using mutual information function, its biological implications and propose a model to account for the observed correlations.

The firsts studies on correlation structure of DNA were reported in 1959 [3], [4]. Almost twenty years later [5]-[9], a burst of interest around the concept of long-range correlation encouraged a fruitful line of research. In [10] a mechanism for the explanation of this property is proposed.

Those studies were inconclusive in the case of human DNA because of the incompleteness of the sequenced data. Human genome is now completely available and we show our first results. The structure of the paper is as follows. In Section II, the mathematical tools and concepts are introduced. Section III we presents the results of our analysis on the 24 human chromosomes. In Sec. IV, the result obtained are discussed.

**II Mutual Information Function.**

Statistical analysis of symbolic sequences as we had said, had a bloom when the first pieces of DNA data were available [5]-[9]. By applying a set of techniques, such as entropies and spectral analysis, the existence of long-range correlation at least in noncoding sectors of DNA was disclosed beyond any reasonable doubt. We would like to stress here that human genome is composed in a 95% of these type of sectors.

The concept of mutual information function can be found for the first time in the seminal 1948 Claude Shannon's paper [11]. His results were generalized to abstract alphabets by several Soviet mathematicians, culminating in the work of R.L. Dobrushin [12]. The original idea was to measure the difference between the average uncertainty in the input of an information channel before and after the output is received. In recent years it has been used to study some properties of regular languages and cellular automata [13], as well as DNA long-range correlations [5]-[9], [14], and some properties of strange attractors [15]. A comprehensive discussion of the properties of this function can be found in [16]-[18]. We use mutual information function to study long-range correlation properties because as proved in [17] it is a more sensitive measure than autocorrelation function.

Let denote by $A = \{A, C, G, T\}$ an alphabet and $s = (\ldots a_0, a_1, \ldots)$ an infinite string with $a_i \in A$, $i \in Z$, where $Z$ represents the set of all integer numbers. The mutual information function of the string $s$ is defined as:

$$M(d,s) = \sum_{\alpha,\beta \in A} P_{\alpha,\beta}(d,s) \ln\left[\frac{P_{\alpha,\beta}(d,s)}{P_\alpha(s) P_\beta(s)}\right] \qquad (3)$$

where the sum is over all pairs $(\alpha, \beta) \in A^2$ and:

$P_{\alpha,\beta}(d,s)$ is the joint probability of having the symbol $\alpha$ followed $d$ sites away by the symbol $\beta$ on the string $s$.

$P_\alpha(s)$ is the density of the symbol $\alpha$ in the string $s$.

Human genome, obviously is not an infinite string, but large enough to guarantee the accuracy of statistics. For instance, chromosome 1 has about 233,819,029 bps.

**III Main results.**

The results of our calculations are shown in Fig. 1. The mutual information function for the 24 human chromosomes are shown there. Although we calculated the values of mutual information function for $d = 1, \ldots, 150,000$ we only show them in the interval $d = 1, \ldots, 1024$. As could be notice, the shape of these curve is the same for all chromosome. The only difference is in the height of the graphs. The order in which these graphs appears from above to below is: 4, 13, Y, X, 6, 5, 18, 3, 8, 2, 7, 12, 21, 14, 9, 10, 11, 1, 20, 15, 16, 17, 22, 19. Every number stands for the chromosome number. The letter X stands for the chromosome X and letter Y for the chromosome Y.

**IV Discussion.**

It seems to us very striking that the mutual information functions of all chromosome have the same shape. The peaks are placed in the same position and only heights are different, reflecting different strength in correlation. It suggests that correlation among bases has the same pattern in all chromosome in spite of the difference in function for different chromosomes. This fact suggest a unique mechanism of structural evolution. In some sense, it is like a novel, in which each chapter develops a different theme, but the correlation among the letters in every chapter is the same.

Another interesting fact is the different slope in different sectors of the curve. Because of the graph is in logarithmic scale, it suggests the existence of more than one exponent for the scaling. In [10] is proved for binary alphabet that a mechanism yielding this behavior is an expansion-modification system [19]. We propose that such a mechanism should account for the behavior of human genome [20].

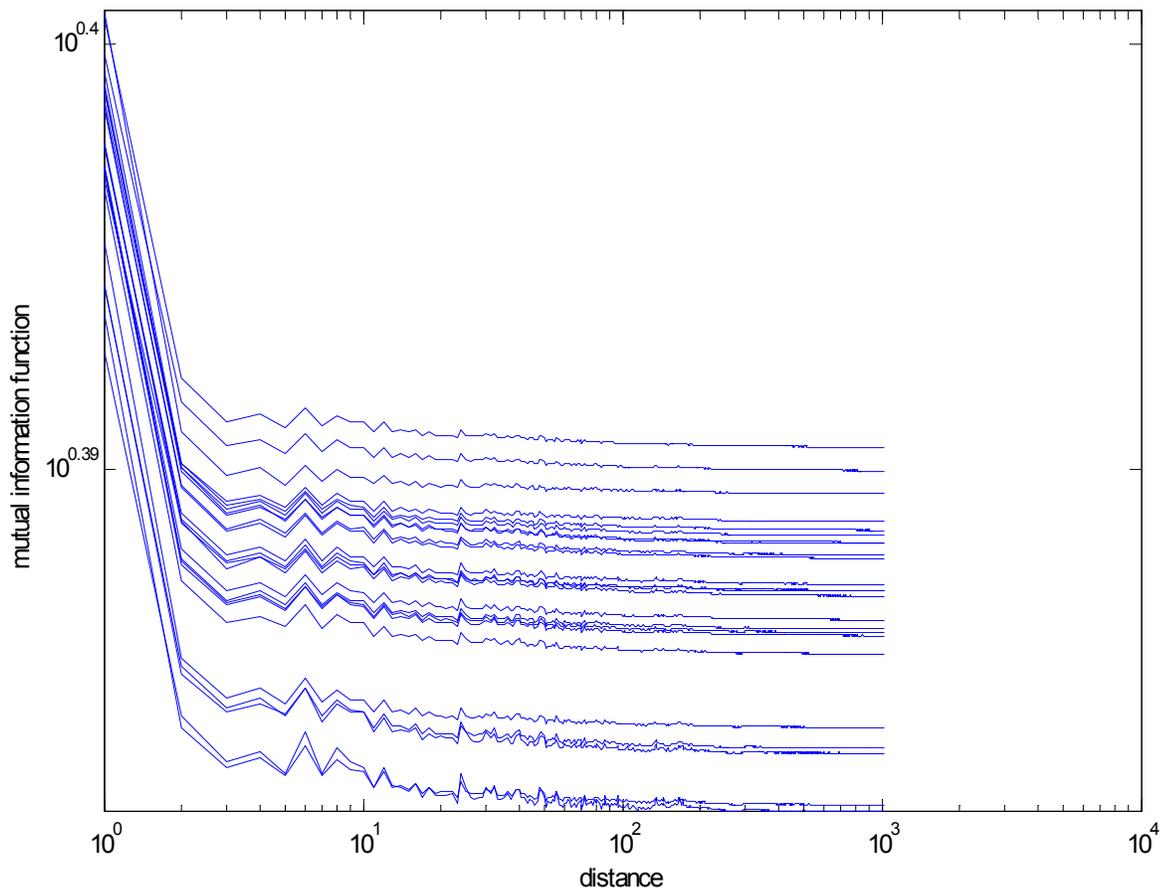

Fig. 1: Graph of the mutual information function for the 24 human chromosome. In the X axis is shown the logarithms of the distances and in the Y axis the logarithms of the functions.